\documentclass[10pt]{article}
\usepackage{graphicx,xcolor,amsmath,amssymb,multirow,ulem,wrapfig,amsfonts,natbib}
\usepackage{epstopdf,epsfig,tikz,amsbsy,mathabx,esint,longtable,subfigure,subcaption,placeins}
\usetikzlibrary{arrows,decorations.markings}
\usetikzlibrary{patterns,positioning,decorations.pathmorphing,arrows,decorations.pathreplacing,calligraphy}
 
\definecolor{OliveGreen}{rgb}{0,0.6,0}
\renewcommand{\Re}{\mathrm{Re}}
\pgfdeclarelayer{back} \pgfsetlayers{back,main}
\definecolor{OliveGreen}{rgb}{0,0.6,0}

\def\XXint#1#2#3{{\setbox0=\hbox{$#1{#2#3}{\int}$}
		\vcenter{\hbox{$#2#3$}}\kern-.5\wd0}}

\makeatletter
\pgfkeys{%
	/tikz/on layer/.code={
		\pgfonlayer{#1}\begingroup
		\aftergroup\endpgfonlayer
		\aftergroup\endgroup
	},
	/tikz/node on layer/.code={
		\pgfonlayer{#1}\begingroup
		\expandafter\def\expandafter\tikz@node@finish\expandafter{\expandafter\endgroup\expandafter\endpgfonlayer\tikz@node@finish}%
	},
}

\usepackage{authblk}

\title{Time-domain pressure and surface wave propagation over generic topography due to sea floor motion}
\author{ R. Pethiyagoda$^1$, $\,\,$   S. Das$^{2,3}$\thanks{Corresponding author: santudas20072@gmail.com, d.santu@iasst.gov.in}, $\,\,$ M. H. Meylan$^1$} 

\date{%
	$^1$School of Information and Physical Sciences, University of Newcastle, NSW 2308, Australia\\%
	$^2$Mathematical and Computational Sciences (Physical Sciences Division), Institute of Advanced Study in Science and Technology, Guwahati 781035, India \\%
	$^3$Academy of Scientific and Innovative Research (AcSIR), Ghaziabad - 201002, India\\[2ex]%
	\today
}

\begin{document}
	\maketitle
	
	\begin{abstract}
		The surface gravity wave evolution, imitating tsunamis triggered by the ocean floor's arbitrary temporal motion over a generic seafloor topography, is investigated using the linearised water wave theory of a compressible ocean. The unprecedented details of pressure wave bounce between the water surface and ocean floor, followed by the generation of acoustic-gravity wave (AGW) propagation away from the initial disturbance, are shown for the first time. The computational novelty covers accurate pressure-wave-field computation using high-frequency AGW modes, unlike the dominant first AGW in surface waves. The mathematical problem is solved using the Fourier transformation and eigenfunction expansion techniques, following a multistep approximation of the arbitrarily deep sea. Test cases of two idealistic subsurface geometries (continental shelf and mountain range) using the ocean floor's linear and parabolic temporal growth show elongation and shortening of the surface wavelength and changes in the propagation speed due to the variable seafloor topography through simulations. The mutual interaction of the pressure waves within the water column unravels depth-change-induced scattering. While high-frequency pressure at a shallower depth is dominant for the continental shelf, a significantly reduced amplitude across a mountain ridge is visualised with a corresponding lower impact of the acoustic-gravity wave on the surface wave. Substantial contrast in the pressure wavefield triggered by a slight change in the rupture's properties suggests its careful consideration when building tsunami prediction models. Since these AGWs are precursors to tsunamis, we believe their detection through an accurate pressure measurement will give very important information about the associated tsunami wave.
	\end{abstract}
	
	\section{Introduction}\label{Section1}
	Tsunami waves generated from submarine earthquakes or landslides encounter numerous topographies of the ocean floor, such as continental shelves, continental slopes, abyssal planes, ocean trenches and plateaus, and mid-oceanic ridges, that lead to the evolution of transoceanic wave propagation \citep{kulikov2016numerical,salaree2020effects}. The pressure wave generated around the initial disturbance also transforms across different topographies and is entangled with the surface profile. The resulting wave evolution and the underwater pressure wave structure quantify the intensity of the tsunami waves as they make landfall along the coastal water, leading to the severity of impending damages \citep{miyashita2022local}.  According to the World Health Organization (WHO), the estimate of deaths caused by tsunamis between $1998-2017$ reached $250000$, with the $2004$ India Ocean Tsunami alone responsible for more than $227000$. The World Bank estimated a total loss of more than US$\$ 7$ billion (http://documents.worldbank.org/curated/en/ 194061468258308532/ World-Bank-response-to-the-Tsunami-disaster). Clearly, the estimation of the tsunami wave profile and pressure wave signature across the ocean, which is influenced by different seafloor topography, is of paramount importance. Here, our objective is to build a mathematical model of submarine earthquake-generated surface waves impacted by arbitrary ocean floor topography in their propagation course, emphasising the computational accuracy of the pressure wave profile that involves contributions from high frequency. Such computational accuracy is lacking in previous calculations, and the present work presents the benchmark computations for pressure wave profiles. 
	
	Given the length scale of tsunami wave propagation, the surface tension and viscosity of ocean water are not as important as the ocean compressibility, which paves the way for a mathematical structure associated with the Acoustic-Gravity Wave (AGW) - a low-frequency hydroacoustic wave influenced by gravity, which plays an important role in accurate computations of the pressure inside the water column. These AGW modes generated in a compressible ocean are analogous to the seismic waves in the Earth. They are compressional waveguide modes in the ocean. We note that the first AGW mode becomes the tsunami wave mode when compression is taken to be infinite. For realistic parameters, this mode is similar to the incompressible tsunami wave. Although the mathematical technique to add ocean water compressibility into the physical system is well established through the works of \cite{longuet1950theory}, \cite{sells1965effect}, \cite{miyoshi1954generation}, \cite{yamamoto1982gravity}, a renewed interest in applying the theory in the tsunami wave propagation model is later found through a series of works by Nosov \citep{nosov1999tsunami,nosov2001nonlinear,nosov2007elastic} which turned out to be talismannic in the aftermath of the devastating Indian Ocean tsunami. These works pointed out the enhanced accuracy of tsunami wave profile estimation primarily due to the impact of AGW in a compressible ocean compared to the same for an incompressible ocean, a common assumption made in the majority of the works. Later, \cite{stiassnie2010tsunamis} and \cite{hendin2013tsunami} provided a solution technique to address the far-field behaviour of the tsunami and AGW modes. With this idea, subsequent researchers have investigated different properties of AGW, including two interacting gravity waves producing \textsc{AGW}\citep {kadri2013generation}, triad resonances \citep{kadri2016triad,kadri2016resonant}, deep-water transport \citep{kadri2014deep}, ice shelf breaking \citep{kadri2012acoustic}, and wave dissipation on sea ice \citep{chen2019dispersion} or influence of different ocean floor characteristics such as porous \citep{chierici2010modeling}, viscous \citep{abdolali2015depth}, or rigid but variable bathymetry using depth-integrated mild-slope equations \citep{sammarco2013depth}, Fourier-series expansion \citep{Das2023time-domain}. However, all of these works neglected one aspect of ocean water compressibility, namely static compression.
	
	Static compression refers to the increase in the density of ocean water as a result of the pressure of the overlying fluid. It is only important for deep ocean water. The associated theory was first developed in the work of \cite{longuet1950theory} but is often neglected. 
	It has only a small significance for typical ocean depths, and inclusion requires an increase in mathematical difficulty. For example, static compression equations appear in \cite{omira2022global} as part of their analysis of tsunami waves generated by the Hunga Tonga–Hunga Ha`apai (HTHH) volcanic eruption, but they then neglect this term in their solution. However, a few recent works \citep{kadri2015wave,abdolali2017role,abdolali2019effect,kadri2016triad,kadri2013generation} have shown its importance in the pure gravity mode (tsunami), although their primary focus was on finding the dispersion relation and the propagation speed. In particular, \cite{abdolali2017role} emphasised the static compression-influenced amplification of the phase speed at a higher water depth. Later,  \cite{meza2023acoustic} introduced a technique to obtain the far-field behaviour of waves based on a simplified dispersion relation associated with a large water depth. The primary issue of dealing with static compression in finite depth is the establishment of an inner product between the complicated depth-dependent functions so that more realistic ocean floor topography can be handled. The issue was addressed by \cite{Das2023PoF}, who provided a complete solution for the canonical problem of vertically moving ocean floor-generated surface wave motion using the eigenfunction expansion method. The simplicity of the method in tackling the mathematical rigour concerning static compression over standard transformation methods makes the approach novel. The method is later extended to three dimensions \citep{das4636389effect} and to model the transformation of surface waves over a step \citep{Das2024} and a trench \citep{Pethiyagoda_Das_Meylan_2024}. Static compression is also introduced to the propagation of wave propagation models from surface disturbances \citep{pethiyagoda4677642atmospheric,pethiyagoda_3d_atmos}, which supplemented the work of \cite{renzi2014hydro} where static compression was ignored and \cite{liu2022water} where an incompressible water was assumed.
	
	In reality, the ocean bed shows features of a non-rigid nature, and prior studies emphasised the impact of elastic \citep{eyov2013progressive,abdolali2019effect,kadri2019effect,williams2023acoustic}, porous \citep{Gomez2023tsunami}, and viscoelastic \citep{bahena2023tsunami} properties. These features lead to dissipation of the AGW modes, especially the elastic nature that contributes to the vanishing of the leading order AGW at shallower depth, where it subsequently gets translated to Scholte and Rayleigh waves \citep{eyov2013progressive,williams2023acoustic} and generation of a very small amplitude surface gravity mode (with cut-off frequency) having propagation speed close to the speed of sound \citep{williams2023acoustic}. In addition, due to the temperature gradient and the salinity of the vertical ocean water profile, water density varies in a one-to-one correspondence with the sound speed within. Recent studies have tried to capture the evolution of AGW propagation in such a system \citep{michele2020effects} and also under Lagrangian formulation \citep{Dubois2024new}. However, we are focused solely on the topographical properties of the seabed, instead of the nature of it and variable ocean density (consequently, a vertically changing sound speed profile), which requires further rigorous mathematical treatment. We build up the solution of the canonical problem in two dimensions of the rigid seabed, focusing explicitly on the computational accuracy of the pressure wave profile and its transient nature entangled with the associated surface gravity wave propagation while solving the full dispersion relation, something overlooked in the majority of the prior research, where the far-field solutions are sought.
	
	While the free-surface elevation is dominated by the first AGW mode, the accurate computation of the water-column pressure wave propagation requires higher AGW modes involving high frequencies, which are computationally difficult. The visualisation of AGW through pressure wave propagation inside the ocean water column has not been achieved until the recent works of the authors of this paper. The present results calculate this wave in unprecedented detail. Detection of waves gives a great deal of information about the underlying system in many branches of physics, not least earthquakes.  Similarly, we believe that detecting these AGWs will give very important information about the associated tsunami wave. Recently, \cite{kadri2025acoustic} attempted a similar problem but focused on the quantum tunnelling effect of AGW, whereas we focus on the effect of the transient initial condition. In this work, we attempt novel computations involving such high-frequency modes for improved pressure wave solutions. Then we are able to extend the solution to tackle arbitrary ocean floor topography using step-type approximations, where the eigenfunction expansion method coupled with the Fourier transformation is used. A few realistic geometries, such as continental slopes and mountain ridges, are taken as test cases. Moreover, a nonlinear temporal growth function associated with ocean floor motion is attempted in addition to a linear one. The manuscript is structured as follows. Section~\ref{sec:formulation} presents the two-dimensional mathematical formulation of the physical problem, followed by the solution method that involves eigenfunction matching across the boundaries between regions in \S~\ref{sec:solution_techniques}. The time domain simulations are detailed in \S~\ref{sec:2d_time_domain}. The manuscript concludes with a brief summary in \S~\ref{sec:conclusion}.
	
	\section{Mathematical formulation and modelling assumptions}\label{sec:formulation}
	
	We briefly discuss the modelling assumptions which underlie this work.  The starting point for our work here is the papers by Stiassnie and coworkers \citep{stiassnie2010tsunamis,hendin2013tsunami}, and almost all of our assumptions follow from those used in this work.  We note these same assumptions are also made in recent work \citep{kadri2025acoustic}. We acknowledge that this work did not include the effect of sea floor elasticity, and neither does the work here.  We note that this effect may be important \citep{abdolali2019effect,williams2023acoustic} in shallow water.  The only modelling change we make to the work of Stiassnie is to include static compression as formulated in the classical work of \citet{sells1965effect}.  In the literature, there has been discussion about the importance of this term.  We include this term primarily to test its effect and because it adds no run-time to the numerical solution. Our results here and in previous works show that its effect is small. We note that more complicated models for static compression have been developed \cite{michele2020effects}, but we anticipate that they will also only have a small effect.  
	Finally, we add a brief note concerning \textit{solving} an equation. We believe that the solution needs to be computed and visualised.  We note that a full numerical solution was not presented in \cite{hendin2013tsunami} and many related works.  We believe that it is essential to capture the entire pressure throughout the fluid, not to simply the motion of the free surface.

\begin{figure}
		\centering
		\begin{tikzpicture}[scale=0.9]
			\draw[fill=gray,opacity=0.7] (-2.5,-4.6) -- (-2.5,-4.3) -- (0.5,-4.3) -- (0.5,-4.6) -- (-2.5,-4.6);
			\draw[fill=black,opacity=0.5] (-8.5,-4.6) -- (-8.5,-3.6) -- (-8,-3.6) -- (-8,-4) -- (-7.6,-4) -- (-7.6, -4.1) -- (-7.2,-4.1) -- (-7.2,-3.9) -- (-6.7,-3.9)-- (-6.7,-3.5) -- (-6,-3.5) -- (-6,-4) -- (-5.9,-4) -- (-5.9,-4.2)-- (-5.6,-4.2) -- (-5.6,-4.3) -- (-4.5,-4.3) -- (-4.5,-4.6) -- (-8.5,-4.6);
			\draw[dashed] (-4.5,-4.3) -- (-4.5, -1.5);
			\draw[dashed] (-5.6,-4.2) -- (-5.6, -1.5);
			\draw[dashed] (-5.9,-4.1) -- (-5.9, -1.5);
			\draw[dashed] (-6,-3.5) -- (-6, -1.5);
			\draw[dashed] (-6.7,-3.5) -- (-6.7, -1.5);
			\draw[dashed] (-7.2,-3.9) -- (-7.2, -1.5);
			\draw[dashed] (-7.6,-4) -- (-7.6, -1.5);
			\draw[dashed] (-8,-3.6) -- (-8, -1.5);
			\draw[fill=black,opacity=0.5] (6.5,-4.6) -- (6.5,-4) -- (6,-4) -- (6,-4.1) -- (5.5,-4.1) -- (5.5,-4.2) -- (5,-4.2) -- (5,-4.05) -- (4.5,-4.05)-- (4.5,-4.15) -- (4,-4.15) -- (4,-4.3) -- (3.5,-4.3) -- (3.5,-4.4)-- (3,-4.4) -- (3,-4.5) -- (2.5,-4.5) -- (2.5,-4.6) -- (6.5,-4.6);
			\draw[dashed] (2.5,-4.5) -- (2.5, -1.5);
			\draw[dashed] (3,-4.5) -- (3, -1.5);
			\draw[dashed] (3.5,-4.4) -- (3.5, -1.5);
			\draw[dashed] (4,-4.3) -- (4, -1.5);
			\draw[dashed] (4.5,-4.15) -- (4.5, -1.5);
			\draw[dashed] (5,-4.05) -- (5, -1.5);
			\draw[dashed] (5.5,-4.2) -- (5.5, -1.5);
			\draw[dashed] (6,-4.1) -- (6, -1.5);
			\fill [cyan, opacity=0.15] (-8.5,-1.4) rectangle (6.5,-4.6); 
			\fill [gray,opacity=0.8] (-8.5,-5.8) rectangle (6.5,-4.6);
			\draw[ultra thick, red] (-3,-1.4) plot[domain=-8*pi:-4*pi, samples=150]  (\x/pi,{sin(\x r)/4-1.4});
			\draw [line width=0.5mm,color=blue,->] (-5,-1) --(-8,-1);
			\draw[ultra thick, red] (3,-1.4) plot[domain=1.5*pi:5.5*pi, samples=150]  (\x/pi,{cos(\x r)/4-1.4});
			\draw (-8.5, -1.2) node{$z=0$};
			\draw (-8.5, -3.4) node{$z=-h_1$};
			\draw (-1.1, -5.5) node{\textbf{Rigid impermeable bottom}};
			\draw [line width=1pt] (-2.5,-4.3)   -- (-2.5,-4.6)node(xline)[below] {$l_{k-1}$};
			\draw [line width=1pt] (-4.5,-4.6)   -- (-4.5,-4.5)node(xline)[below] {$l_{k-2}$};
			\draw [line width=1pt] (0.5,-4.3) --(0.5,-4.6) node (yline)[below]{$l_k$};
			\draw [line width=1pt] (2.5,-4.5) --(2.5,-4.6) node (yline)[below]{$l_{k+1}$};
			\draw [line width=1pt] (-8.5,-4.6)node (yline)[below]{$l_0$} --(6.5,-4.6)node (yline)[below]{$l_{m_1}$};
			\draw[line width=0.5mm,color=black] (-8.5,-1.4) -- node[below] {} ++(5.5,0) -- node[below] {} ++(4,0) -- (6.5,-1.4);
			\draw (-2.5,-2) node{$\displaystyle \nabla^2\Phi=\frac{1}{c^2}\left(\Phi_{tt} +g\Phi_z\right)$};
			\draw[color=red] (-1,-3.3) node {$\mathcal{R}_k$};
			\draw[color=red,dash dot,->] (-0.7,-3.3) -- (0.45,-3.3);
			\draw[color=red,dash dot,->] (-1.3,-3.3) -- (-2.5,-3.3);
			\draw[color=red, dashed] (0.5,-2.6) -- (0.5,-4);
			\draw[color=red] (1.5,-3.3) node {$\mathcal{R}_{k+1}$};
			\draw[color=red,dash dot,->] (2,-3.3) -- (2.4,-3.3);
			\draw[color=red,dash dot,->] (1.1,-3.3) -- (0.55,-3.3);
			\draw[color=red, dashed] (-2.52,-2.6) -- (-2.52,-4);
			\draw[color=red] (-3.5,-3.3) node {$\mathcal{R}_{k-1}$};
			\draw[color=red,dash dot,->] (-3,-3.3) -- (-2.55,-3.3);
			\draw[color=red,dash dot,->] (-4,-3.3) -- (-4.4,-3.3);
			\draw [line width=1pt,->] (-1,-1.25)  -- (0,-1.25) node(xline)[right] {$x$};
			\draw [line width=1pt,->] (-1,-1.25) -- (-1,-0.25) node(yline)[above] {$z$};
			\draw[line width=0.5mm,color=blue,->] (2.5,-1)-- (5.5,-1);
			\draw[line width=0.5mm,color=blue,->] (-1.75,-4.6)-- (-1.75,-4.3);
			\draw [line width=0.5mm,color=blue,->] (-0.25,-4.6) --(-0.25,-4.3);
			\draw [ultra thick,decorate, decoration = {brace}] (0.58,-4.3) --  (0.58,-4.6);
			\draw (0.65,-4.5)node(yline)[right] {$a_0$};
		\end{tikzpicture}
		\caption{Schematic of the physical problem depicting the ocean floor's vertical movement. which is amplified for visual clarity. The block width and ocean depth have length scales of kilometers, whereas $a_0$ is in meters. In addition, $a_0\ll(h_i-h_j)$ (for $i,j=0,1,\dots, m_1$) is imposed to show the significant local ocean depth.}
		\label{Fig1}
	\end{figure}
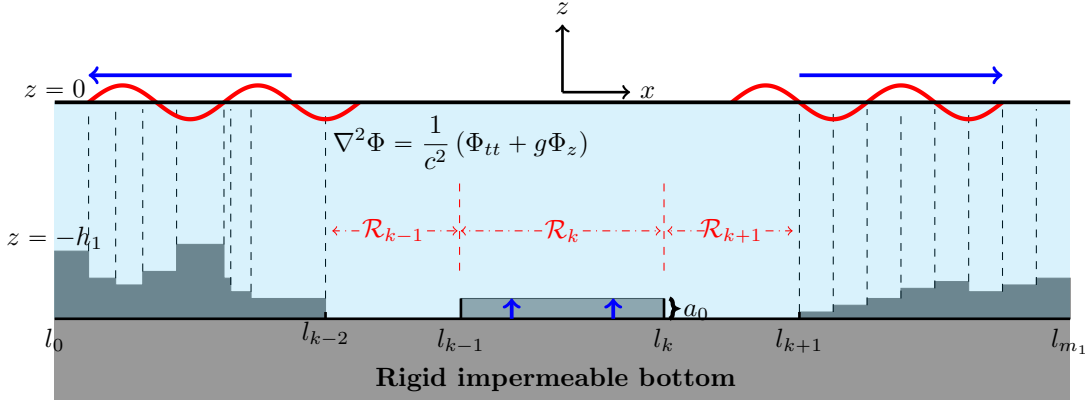

	The generation of free-surface gravity waves from the finite-time arbitrary vertical movement of a block of the ocean bottom of fixed length in a compressible ocean, including the effect of static background compression, is considered. The vertical movement has a maximum uplift of $a_0$ and a duration of $\tau$ seconds, which induces a rise in the free surface and results in wave propagation. The ocean bathymetry is arbitrary and approximated with step-type topography at the locations $x=l_i\,(i=0,1,\dots,m_1)$ with assumptions $l_0\to -\infty$ and $l_{m_1}\to\infty$ and local ocean depth $h_i$ in the region $(l_{i-1},\,l_i)$ (say region $\mathcal{R}_i$). Without loss of generality, we assume that the movement takes place in $\mathcal{R}_k$. (Fig. \ref{Fig1}).  A two-dimensional Cartesian coordinate system, centred in the middle of the movement, is considered for a mathematical formulation involving the $z$-axis being perpendicular to the horizontal $x$-axis and pointing upwards. Under the assumption of irrotational flow, we solve the boundary value problem (\textsc{BVP}) constructed with the velocity potential $\Phi(x,z,t)$ inside the fluid domain that takes different forms based on the ocean depth. Assuming a ridge impermeable bottom, the fluid velocity in the vertical direction is given by the bathymetry velocity, $\mathcal{I}(x,t)$.
	
	The boundary value problem (BVP) inside the trench in the time domain is given by
	\allowdisplaybreaks
	\begin{subequations}\label{BVP_time_in}
		\begin{align}
			\Phi^{(i)}_{xx}+\Phi^{(i)}_{zz}&= \frac{1}{c^2}\left(\Phi^{(i)}_{tt}+g \Phi^{(i)}_{z}\right),\quad &-h_i<z<0,\,x\in \mathcal{R}_i,\\
			\Phi^{(i)}_{tt}+g\Phi^{(i)}_{z}&=0, \quad &z=0,\,x\in \mathcal{R}_i,\\
			\Phi^{(i)}_{z}&=\mathcal{I}(x,t), &z=-h_i,\,x\in \mathcal{R}_i,
		\end{align} 
	\end{subequations}
	where $i=0,\,1,\dots,m_1$, $c$ represents the speed of sound in the water, and the vertical velocity along the bottom is assumed separable and given by
	\begin{align*}
		\mathcal{I}(x,t)=\left\{\begin{matrix}
			\mathcal{L}(t) & \mbox{when }x\in \mathcal{R}_k,\\
			0 & \mbox{when }x\notin \mathcal{R}_k
		\end{matrix}
		\right.
	\end{align*}
	where $\mathcal{L}(t)$ is the temporal velocity of the ocean floor. Note that the term $g\phi_z^{(i)}$ in Eqs. (\ref{BVP_time_in}a,b) is significant for surface-gravity waves as opposed to the AGW \citep{michele2020effects}. However, we are not interested here in the individual impact of either the AGW modes or the surface gravity mode, but rather the combined effect of both types on the pressure wave propagation, focusing on accurately capturing the qualitative behaviour using higher-order modes. While a decoupling of the IBVP based on the mode types (i.e., with or without the terms $g\phi_z^{(i)}$) seems the logical choice, we choose to retain the term in the subsequent solution process to maintain a unified setup.
	
	The matching conditions at the edge of depth change are specified as
		\begin{align}
			\left(\begin{matrix}\label{match_time}
				\Phi^{(i)}\\
				\Phi^{(i)}_x
			\end{matrix} \right)
			= \left(\begin{matrix}
				\Phi^{(i+1)}\\
				\Phi^{(i+1)}_x
			\end{matrix} \right)
			\quad\mbox{at} \quad x=l_i,-h_{i,m}<z<0\quad \text{for }m=1,\,2,\dots,m_1-1,
		\end{align}
	where $h_{i,m}=\text{min} (h_i,\,h_{i+1})$, with the no-flux condition
		\begin{align}\label{noflux_time}
			\left.\begin{matrix}
				\partial_x\Phi^{(i+1)}=0 & \mbox{when }h_i<h_{i+1}\\
				\partial_x\Phi^{(i)}=0 & \mbox{when }h_i>h_{i+1}
			\end{matrix}\right\}&\mbox{at } x= l_i,\,-h_{i,M}<z<-h_{i,m}\quad \text{for }i=1,\,2,\dots,m_1-1.
		\end{align}
		where $h_{i,M}=\text{max} (h_i,\,h_{i+1})$.
		
		\section{Solution technique}\label{sec:solution_techniques}
		Applying the Fourier transformation in time of the form 
		\begin{subequations}
			\begin{align}
				\phi(x,z,\omega)=\int_{-\infty}^{ \infty} \Phi(x,z,t)e^{-{\rm i}\omega t}dt,
			\end{align}
			whose inverse transformation is
			\begin{align}\label{inverse}
				\Phi(x,z,t)=\frac{1}{2\pi}\int_{-\infty}^{\infty} \phi(x,z,\omega)e^{ {\rm i}\omega t}d\omega.
			\end{align}
		\end{subequations}
		Note that in our problem, the solution will be zero for $t<0$. This makes the solution here very close to a Laplace transform solution with the variable $s$ replaced by $-\mathrm{i}\omega$ and the Bromwich inversion contour on the imaginary ($s$ axis). We note that the Fourier transform is generally more numerically stable than the Laplace transform. 
		
		We transform the BVPs, matching conditions, and no-flux conditions stated in Eqs. \eqref{BVP_time_in} -- \eqref{noflux_time} into the frequency domain BVP in the inner region as
		\allowdisplaybreaks
		\begin{subequations}\label{BVP_freq_in}
			\begin{align}
				\left(\partial_{xx}+\partial_{zz}  \right)\phi^{(i)}&= \frac{1}{c^2}\left(-\omega^2+g\partial_{z}\right)\phi^{(i) },\quad &h_i<z<0,\,x\in \mathcal{R}_i,\\
				\left(-\omega^2+g\partial_{z}\right) \phi^{(i)}&= 0,\quad &z=0,\,x\in \mathcal{R}_i,\\
				\partial_z\phi^{(i)}&=\xi(\omega), \quad &z=-h_i,\,x\in \mathcal{R}_i,
			\end{align} 
		\end{subequations}
		where 
		\begin{align*}
			\xi(\omega)=\left\{\begin{matrix}
				\displaystyle\int_{-\infty}^{ \infty} \mathcal{L}(t)e^{-{\rm i}\omega t}dt & \mbox{when }x\in \mathcal{R}_k,\\
				0 & \mbox{when }x\notin \mathcal{R}_k.
			\end{matrix}
			\right.
		\end{align*}
		The matching conditions at $x=l_i$ are transformed into
		\begin{align}
			\left(\begin{matrix}\label{match_time}
				\phi^{(i)}\\
				\phi^{(i)}_x
			\end{matrix} \right)
			= \left(\begin{matrix}
				\phi^{(i+1)}\\
				\phi^{(i+1)}_x
			\end{matrix} \right)
			&\quad\mbox{along} \quad h_{i,m}<z<0,\\
			\left.\begin{matrix}
				\partial_x\phi^{(i+1)}=0 & \mbox{when }h_i<h_{i+1}\\
				\partial_x\phi^{(i)}=0 & \mbox{when }h_i>h_{i+1}
			\end{matrix}\right\}&\quad \mbox{along}\quad -h_{i,M}<z<-h_{i,m}.
		\end{align}
		Letting $\varphi(x,z,\omega)$ be the corresponding potential function such that 
		\begin{align}\label{inter_pot}
			\phi^{(i)}(x,z,\omega)=\xi(\omega) \varphi^{(i)}(x,z,\omega),
		\end{align}
		we can write the potential as
		\begin{subequations}\label{pot_freq}
			\begin{align}
				\varphi^{(i)}&=\sum_{n=0}^{\infty} \left\{b_{i,n}e^{{\rm i}k_n^{(i)} (x-l_{i-1})} + c_{i,n}e^{-{\rm i}k_n^{(i)} (x-l_i)} \right\}f_n^{(i)}(z)+f^{p(i)}(z),\quad \mbox{for}\quad i=1,\ldots,m_1,
			\end{align}
		\end{subequations}
		with $b_{1,n}=0$, $c_{m_1,n}=0$ for all $n$. The depth-dependent functions are expressed as (following \cite{Das2023PoF})
		\begin{subequations}
			\begin{align}
				f_n^{(i)}(z)=&\frac{\displaystyle e^{(\gamma z/2)}\left( \frac{\gamma}{2\mu^{(i)}_n}\sinh{\mu^{(i)}_n(z+h_i)}-\cosh{\mu^{(i)}_n(z+h_i)} \right)}{\frac{\gamma}{2\mu^{(i)}_n}\sinh{( \mu^{(i)}_n h_i)}-\cosh{(\mu^{(i)}_n h_i)}}\,\mbox{for }i=1,2,\dots,m_1,\\
				f^{p(i)}(z)=&\left\{\begin{matrix}
					\displaystyle\frac{\displaystyle e^{\frac{\gamma (z+h_1)}{2}}\left[\left( \frac{\omega^2}{gk_s}-\frac{\gamma}{2k_s} \right)\sinh{(k_sz)} +\cosh{(k_sz)} \right]}{\displaystyle\left( \frac{\omega^2}{c^2k_s} - \frac{\omega^2\gamma}{2gk_s} \right) \sinh{(k_sh_1)}+\frac{\omega^2}{g}\cosh{(k_sh_i)}} & \mbox{for }i=k,\\
					0 & \mbox{for }i\neq k,
				\end{matrix}
				\right.
			\end{align}
		\end{subequations}
		with $\gamma=g/c^2$ and $\mu^{(i)}_n$ satisfying the following dispersion relation
		\begin{align}
			\frac{\omega^2}{g}=\mu^{(i)}_n \frac{\displaystyle\left[1- \left( \frac{\gamma}{2\mu^{(i)}_n} \right)^2 \right]\tanh{\mu^{(i)}_n h_i}}{\displaystyle 1-\left(\frac{ \gamma}{2\mu^{(i)}_n} \right)\tanh{\mu^{(i)}_n h_i}}\quad\mbox{for }i=1,2,\dots,m_1,
		\end{align}
		and 
		\begin{equation*}
			k_s^2=\frac{\omega^2}{c^2}-\left(\frac{\gamma}{2}\right)^2\quad\mbox{and} \quad (k^{(i)}_n)^2=\frac{\omega^2}{c^2} -\left(\frac{\gamma}{2}\right)^2+(\mu^{(i)}_n)^2.
		\end{equation*}
		
		Applying matching conditions at $x=l_i$ $(i=1,2,\dots, m_1)$ yield
		\begin{subequations}\label{eq:matching}
			\begin{align}
				\sum_{n=0}^{\infty} \left\{b_{i,n}e^{{\rm i}k_n^{(i)} (l_i-l_{i-1})} + c_{i,n} \right\}f_n^{(i)}(z)+f^{p(i)}(z) &= \sum_{n=0}^{\infty} \left\{b_{i+1,n}+ c_{i+1,n}e^{-{\rm i}k_n^{(i+1)} (l_i-l_{i+1})} \right\}\nonumber\\
				&\qquad\quad\times f_n^{(i+1)}(z)+f^{p(i+1)}(z),\\
				\sum_{n=0}^{\infty} k_n^{(i)}\left\{b_{i,n}e^{{\rm i}k_n^{(i)} (l_i-l_{i-1})} - c_{i,n} \right\}f_n^{(i)}(z) &= \sum_{n=0}^{\infty}k_n^{(i+1)} \left\{b_{i+1,n}- c_{i+1,n}e^{-{\rm i}k_n^{(i+1)} (l_i-l_{i+1})} \right\}\nonumber\\
				&\qquad\quad\times f_n^{(i+1)}(z)
			\end{align}
		\end{subequations}
		Now, we construct the system of equations.
		
		\subsection{Construction of the system of equations}
		To construct a system of equations to obtain the coefficients $b_{i,n}$, and $c_{i,n}$ after truncating the series up to $n=N$, we take the inner product defined as
		\begin{align}
			\left<\mathcal{F}_i(z),\, \mathcal{F}_{i+1}(z) \right>_{c}^{h}:= \int_{-h}^0{e^{-\gamma z}\mathcal{F}_i(z)\mathcal{F}_{i+1}(z) \,dz}\quad\mbox{for}\quad i=1,2,\dots,m_1,
		\end{align}
		of equations (\ref{eq:matching}) with $\mathcal{F}_i=f_n^{(i)}$ and $h=h_{i,m}$ or $h_{i,M}$. Specifically, we use $<,>_{c}^{h_{i,m}}$, and $<,>_{c}^{h_{i,M}}$ respectively in Eq. \eqref{eq:matching}(a--b). Using the orthogonality of either $f_n^{(i)}(z)$ or $f_n^{(i+1)}(z)$, we get the following system of equations for each $j=0,1,\dots,N$.
		\begin{subequations}\label{eq_mount}
			\begin{align}
				\sum_{n=0}^{N}&\left\{e^{{\rm i}k_n^{(i)}(l_i-l_{i-1})} b_{i,n}+ c_{i,n}\right\}\left< f^{(i)}_n,\, f^{(i+1)}_j \right>_{c}^{h_{i,m}}+ \left<f^{p(i)}, \,f^{(i+1)}_j\right>_{c }^{h_{i,m}}\nonumber\\
				&=\left\{b_{i+1,j}+c_{i+1,j}e^{-{\rm i}k_j^{(i+1)}(l_i-l_{i+1})} \right\}\left< f^{(i+1)}_j,\, f^{(i+1)}_j\right>_{c }^{h_{i,m}}+ \left<f^{p(i+1)}, \,f^{(i+1)}_j\right>_{c }^{h_{i,m}}
			\end{align}
			\begin{align}
				k_j^{(i)}&\left\{e^{{\rm i}k_j^{(i)}(l_i-l_{i-1})} b_{i,j}- c_{i,j} \right\}\left< f^{(i)}_j,\, f^{(i)}_j \right>_{c}^{h_{i,M}}\nonumber\\
				&=\sum_{n=0}^{N} k_n^{(i+1)}\left\{b_{i+1,n}-c_{i+1,n}e^{-{\rm i}k_n^{(i+1)}(l_i-l_{i+1})} \right\}\left< f^{(i+1)}_n ,\, f^{(i)}_j\right>_{c }^{h_{i,M}},
			\end{align}
			when $h_{i+1}<h_i$, or
			\begin{align}
				&\left\{e^{{\rm i}k_j^{(i)}(l_i-l_{i-1})} b_{i,j}+ c_{i,j}\right\}\left< f^{(i)}_j,\, f^{(i)}_j \right>_{c}^{h_{i,m}}+ \left<f^{p(i)}, \,f^{(i)}_j\right>_{c }^{h_{i,m}}
				\nonumber\\
				&\qquad =\,\sum_{n=0}^{N}\left\{b_{i+1,n}+c_{i+1,n}e^{-{\rm i}k_n^{(i+1)}(l_i-l_{i+1})} \right\}\left< f^{(i+1)}_n,\, f^{(i)}_j\right>_{c }^{h_{i,m}}+ \left<f^{p(i+1)}, \,f^{(i)}_j\right>_{c }^{h_{i,m}},
			\end{align}
			
			\begin{align}
				\sum_{n=0}^{N}&k_n^{(i)} \left\{e^{{\rm i}k_n^{(i)}(l_i-l_{i-1})} b_{i,n}- c_{i,n}\right\}\left< f^{(i)}_n,\, f^{(i+1)}_j \right>_{c}^{h_{i,M}}
				\nonumber\\
				&\qquad=\,k_j^{(i+1)}\left\{b_{i+1,j}-c_{i+1,j}e^{-{\rm i}k_j^{(i+1)}(l_i-l_{i+1})} \right\}\left< f^{(i+1)}_j,\, f^{(i+1)}_j\right>_{c }^{h_{i,M}},
			\end{align}
		\end{subequations}
		when $h_{i+1}>h_i$, where the inner products are given by
		\begin{subequations}\label{eq:inner_partiular}
			\begin{align}
				\left<f^{(i)}_n,\,f^{(i)}_j \right>_c^{h_i}=&\frac{\displaystyle\frac{ \sin{2\mu^{(i)}_nh_i}}{4\mu^{(i)}_n}\left(1- \frac{\gamma^2}{4(\mu^{(i)}_n)^2} \right)+\frac{h_i}{2}\left(1+\frac{ \gamma^2}{4(\mu^{(i)}_n)^2}\right)+\frac{\gamma\left( \cos{2\mu^{(i)}_nh_i} -1\right)}{4(\mu^{ (i)}_n)^2}}{\left[\frac{\gamma}{2\mu^{(i)}_n} \sin{(\mu^{(i)}_n h_i)}-\cos{(\mu^{ (i)}_n h_i)} \right]^2} \delta_{jn},\\
				\left<f^{(1)}_j,\,f^p \right>_c^{h_1}=&\frac{f^{(1)}_j(-h_1)e^{\gamma h_1}}{{\mu^{(1)}_j}^2-k_s^2},\\
				\left<f^{(i)}_j,\,f^{(i+1)}_n\right >_{c}^{h}= &\frac{f^{(i)}_j(-h){f^{(i+1)}_n}'(-h)e^{\gamma h}}{{\mu^{(i)}_j}^2-{\mu^{(i+1)}_n}^2},
			\end{align}
		\end{subequations}
		where $\delta_{mn}$ is the Kronecker delta. After scaling by the repeated inner product, $\left< f^{(i)}_j,\, f^{(i)}_j \right>_{c}^{h}$ or $\left< f^{(i+1)}_j,\, f^{(i+1)}_j \right>_{c}^{h}$, we recover the linear subproblem
		\begin{equation}
			\left[\begin{matrix}
				N_{i1} & N_{i2} & N_{i3} & N_{i4}\\
				N_{i5} & N_{i6} & N_{i7} & N_{i8}
			\end{matrix}\right]
			\left[\begin{matrix}
				B_{i} \\C_{i}\\B_{i+1} \\C_{i+1}
			\end{matrix}\right]
			=
			\left[\begin{matrix}
				\hat{d}_i\\
				0_d
			\end{matrix}\right],
		\end{equation}
		where the $(j,n)$th element of each $N_{i\kappa}$ is defined as 
		\allowdisplaybreaks
		\begin{align*}
			N_{i1}&=\left\{
			\begin{matrix}
				\displaystyle\left[e^{{\rm i}k_n^{(i)}(l_i-l_{i-1})}\frac{\left< f^{(i)}_n,\, f^{(i+1)}_j \right>_{c}^{h_{i,m}}}{\left< f^{(i+1)}_j,\, f^{(i+1)}_j\right>_{c }^{h_{i,m}}}\right]_{jn} & \mbox{when } h_{i+1}<h_i,\\
				\mbox{diag} \left(e^{ {\rm i} k_j^{(1)}(l_i-l_{i-1})} \right)_{1\times (N+1)} & \mbox{when } h_{i+1}>h_i,
			\end{matrix}
			\right.\\
			N_{i2}&=\left\{
			\begin{matrix}
				\displaystyle\left[\frac{\left< f^{(i)}_n,\, f^{(i+1)}_j \right>_{c}^{h_{i,m}}}{\left< f^{(i+1)}_j,\, f^{(i+1)}_j\right>_{c }^{h_{i,m}}}\right]_{jn} & \mbox{when } h_{i+1}<h_i,\\
				\mbox{diag}\left\{(1)_{1 \times (N+1)}\right\} & \mbox{when } h_{i+1}>h_i,
			\end{matrix}
			\right.\\
			N_{i3}&=\left\{
			\begin{matrix}
				-\mbox{diag}\left\{(1)_{1 \times (N+1)}\right\} & \mbox{when } h_{i+1}<h_i,\\
				\displaystyle-\left[\frac{\left< f^{(i+1)}_n,\, f^{(i)}_j \right>_{c}^{h_{i,m}}}{\left< f^{(i)}_j,\, f^{(i)}_j\right>_{c }^{h_{i,m}}}\right]_{jn} & \mbox{when } h_{i+1}>h_i,
			\end{matrix}
			\right.\\
			N_{i4}&=\left\{
			\begin{matrix}
				-\mbox{diag} \left(e^{-{\rm i} k_j^{(i+1)}(l_i-l_{i+1})} \right)_{1\times (N+1)} & \mbox{when } h_{i+1}<h_i,\\
				-\displaystyle\left[e^{-{\rm i}k_n^{(i+1)}(l_i-l_{i+1})} \frac{\left< f^{(i+1)}_n,\, f^{(i)}_j \right>_{c}^{h_{i,m}}}{\left<f^{(i)}_j,\, f^{(i)}_j\right>_{c }^{h_{i,m}}}\right]_{jn} & \mbox{when } h_{i+1}>h_i,
			\end{matrix}
			\right.\\
			N_{i5}&=\left\{
			\begin{matrix}
				\mbox{diag} \left(e^{ {\rm i} k_j^{(i)}(l_i-l_{i-1})} \right)_{1\times (N+1)} & \mbox{when } h_{i+1}<h_i,\\
				\displaystyle\left[\frac{ k_n^{(i)}e^{{\rm i}k_n^{(i)}(l_i-l_{i-1})}}{k_j^{(i+1)}} \frac{\left< f^{(i)}_n,\, f^{(i+1)}_j \right>_{c}^{h_{i,M}}}{\left< f^{(i+1)}_j,\, f^{(i+1)}_j\right>_{c }^{h_{i,M}}}\right]_{jn} & \mbox{when } h_{i+1}>h_i,
			\end{matrix}
			\right.\\
			N_{i6}&=\left\{
			\begin{matrix}
				-\mbox{diag}\left\{(1)_{1 \times (N+1)}\right\} & \mbox{when } h_{i+1}<h_i,\\
				\displaystyle-\left[\frac{ k_n^{(i)}}{k_j^{(i+1)}} \frac{\left< f^{(i)}_n,\, f^{(i+1)}_j \right>_{c}^{h_{i,M}}}{\left< f^{(i+1)}_j,\, f^{(i+1)}_j\right>_{c }^{h_{i,M}}}\right]_{jn} & \mbox{when } h_{i+1}>h_i,
			\end{matrix}
			\right.\\
			N_{i7}&=\left\{
			\begin{matrix}
				-\displaystyle\left[\frac{ k_n^{(i+1)}}{k_j^{(i)}} \frac{\left< f^{(i+1)}_n,\, f^{(i)}_j \right>_{c}^{h_{i,M}}}{\left< f^{(i)}_j,\, f^{(i)}_j \right>_{c }^{h_{i,M}}} \right]_{jn} & \mbox{when } h_{i+1}<h_i,\\
				-\mbox{diag}\left\{(1)_{1 \times (N+1)}\right\}  & \mbox{when } h_{i+1}>h_i,
			\end{matrix}
			\right.\\
			N_{i8}&=\left\{
			\begin{matrix}
				\displaystyle\left[\frac{ k_n^{(i+1)}e^{-{\rm i}k_n^{(i+1)}(l_i-l_{i+1})}}{k_j^{(i)}} \frac{\left< f^{(i+1)}_n,\, f^{(i)}_j \right>_{c}^{h_{i,M}}}{\left< f^{(i)}_j,\, f^{(i)}_j \right>_{c }^{h_{i,M}}} \right]_{jn} & \mbox{when } h_{i+1}<h_i,\\
				\mbox{diag} \left(e^{-{\rm i} k_j^{(i+1)}(l_i-l_{i+1})} \right)_{1\times (N+1)}  & \mbox{when } h_{i+1}>h_i,
			\end{matrix}
			\right.
		\end{align*}
		for $j,n=0,1,\dots,N$, and $B_{i}=[b_{i,0},b_{i,1},\dots, b_{i,N}]^T$, $C_{i}=[c_{i,0},c_{i,1},\dots,c_{i,N}]^T$ are column vectors of length $N+1$ containing unknowns, $\hat{d}_{i}=[d_{i,0},d_{i,1},\dots,d_{i,N}]^T$ is a known column vector of length $N+1$ defined by
		\begin{align*}
			\hat{d}_{i,j}=\begin{cases}
				\displaystyle\frac{\left<f^{p(i+1)}, \, f^{(i+1)}_j\right>_{c }^{h_{i,m}}-\left<f^{p(i)}, \, f^{(i+1)}_j\right>_{c }^{h_{i,m}}}{\left< f^{(i+1)}_j,\, f^{(i+1)}_j \right>_{c }^{h_{i,m}}}&h_{i+1}<h_i\\\displaystyle\frac{\left<f^{p(i+1)}, \,f^{(i)}_j \right>_{c }^{h_{i,m}}-\left<f^{p(i)}, \,f^{(i)}_j \right>_{c }^{h_{i,m}}}{\left< f^{(i)}_j,\, f^{(i)}_j \right>_{c }^{h_{i,m}}}&h_{i+1}>h_i
			\end{cases}
		\end{align*}
		for $j=0,1,\dots,N$, and $0_d$ is a column vector of zeros of length $N+1$. We combine the subproblem for every interface into a single linear system, recalling that $b_{1,n}=0$, $c_{m_1,n}=0$ for all $n$,
		\begin{align}\label{matrix_mount}
			&\left[\begin{array}{cccccccccccc}
				N_{1\,2} & N_{1\,3} & N_{1\,4} & \cdots & 0_A & 0_A & 0_A & 0_A & \cdots & 0_A & 0_A & 0_A\\
				N_{1\,6} & N_{1\,7} & N_{1\,8} & \cdots & 0_A & 0_A & 0_A & 0_A & \cdots & 0_A & 0_A & 0_A\\
				\vdots & \vdots & \vdots & \ddots & \vdots & \vdots & \vdots & \vdots & \ddots & \vdots & \vdots & \vdots\\
				0_A & 0_A & 0_A & \cdots & N_{i1} & N_{i2} & N_{i3} & N_{i4} & \cdots & 0_A & 0_A & 0_A\\
				0_A & 0_A & 0_A & \cdots & N_{i5} & N_{i6} & N_{i7} & N_{i8} & \cdots & 0_A & 0_A & 0_A\\
				\vdots & \vdots & \vdots & \ddots & \vdots & \vdots & \vdots & \vdots & \ddots & \vdots & \vdots & \vdots\\
				0_A & 0_A & 0_A & \cdots & 0_A & 0_A & 0_A & 0_A & \cdots & N_{m_1-1\,1} & N_{m_1-1\,2} & N_{m_1-1\,3} \\
				0_A & 0_A & 0_A & \cdots & 0_A & 0_A & 0_A & 0_A & \cdots & N_{m_1-1\,5} & N_{m_1-1\,6} & N_{m_1-1\,7}
			\end{array}\right]\nonumber\\
			&\qquad\times\left[
			\begin{array}{cccccccccccc}
				C_{1} & B_{2} & C_{2} & \cdots & B_{i}  & C_{i} & B_{i+1} & C_{i+1} & \cdots & B_{m_1-1} & C_{m_1-1} & B_{m_1}
			\end{array}\right]^{T} \nonumber\\
			&\qquad=\left[\begin{array}{cccccccccccc}
				\hat{d}_1 & 0_d & \cdots & \hat{d}_i &  0_d & \cdots & \hat{d}_{m_1-1} & 0_d
			\end{array}\right]^{T},
		\end{align}
		where each $0_A$ is a zero square matrix of order $N+1$.  We solve the Eq. \eqref{matrix_mount} to calculate the unknown coefficient and evaluate the potential functions in Eq. \eqref{pot_freq}. 
		
		Once we get the coefficients, we use Eq. \eqref{inter_pot} and then apply the inverse Fourier transformation to obtain the potential functions and, hence, the surface profile and the pressure distribution inside the water column in the time domain.
		
		\section{Time-domain simulation}\label{sec:2d_time_domain}
		The potential functions in the time domain are obtained by combining Eqs. \eqref{inverse}, \eqref{inter_pot} and \eqref{pot_freq} as
		\begin{align}\label{pot_time}
			\Phi^{(i)}(x,z,t)=\frac{1}{2 \pi} \int_{-\infty}^{\infty}\xi(\omega) &\left( \sum_{n=0}^{\infty}\{ b_{i,n} e^{{\rm i}k_n^{(i)} (|x|-l_{i-1})} + c_{i,n} e^{-{\rm i}k_n^{(i)} (|x|-l_i)}\}f_n^{(i)}(z) + f^{p(i)}(z) \right)\nonumber\\
			&\times e^{ {\rm i}\omega t}d\omega.
		\end{align}
		The free-surface elevation is computed as
		\begin{equation}\label{surf_time}
			\eta^{(i)}=-\frac{1}{g}\left.\frac{ \partial\Phi^{(i)}}{\partial t}\right|_{z=0}=-\frac{{\rm i}}{2 \pi g}\int_{-\infty}^{\infty}\omega\phi^{(i)}(x,0,\omega)e^{ {\rm i}\omega t}\,d\omega
		\end{equation}
		for $i=1,2,\dots,N$, where $\Re(\cdot)$ represents the real part of a complex quantity. The pressure distribution inside the water column is computed as
		\begin{equation}\label{press_time1}
			P^{(i)}=-\rho\frac{ \partial\Phi^{(i)}}{\partial t}=-\frac{{\rm i}\rho}{2 \pi}\int_{-\infty}^{\infty} \omega \phi^{(i)}(x,z,\omega)e^{ {\rm i}\omega t}\,d\omega
		\end{equation}
		All integrals in Eqs. (\ref{pot_time})--(\ref{press_time1}) have simple poles along the real axis corresponding to when the determinant of the coefficient matrix in Eq. (\ref{matrix_mount}) is zero. The contour of integration passes over the poles for $\omega<0$ and under for $\omega>0$. The integral is evaluated following \citet{Pethiyagoda_Das_Meylan_2024}, whereby the integral over the truncated domain $\omega\in[0,\omega_\mathrm{max}]$ is subdivided in regions near the poles and elsewhere. The regions near the poles are further modified by adding ans subtracting the singularity resulting in continuous integral and a closed form evaluation of a Cauchy principal valued integral and half-residue contribution. All remaining integrals are evaluated numerically using a Filon-type quadrature.
		
		Now, we consider a few specific bathymetry profiles and two different velocity profiles for bottom rupture, constant velocity $\mathcal{L}(t)=\displaystyle \frac{a_0}{\tau}H(\tau-t)H(t)$ and parabolic velocity $\mathcal{L}(t)=\displaystyle \frac{6a_0}{\tau}t(\tau-t)H(\tau-t)H(t)$. The fixed parameter values are $a_0=1$ m,$c=1450$ m/s, $g=9.8$ m/s$^2$, $\tau=10$ s, and $N=30$, unless otherwise mentioned. Our choice of the topography parameters in the following subsections is idealistic, and is utilised to show the robustness of the method. More accurate topography from the observational data, something we are lacking in collecting at present, may be fit using a proper step-approximation to validate the solution with observational data. Nevertheless, we move forward with the idealistic parameters.  For the calculations here, there are no solutions in the literature to validate against, except our previous solutions for the constant depth case (which were validated against previous results in the literature). We have confidence in our results because we can check that the frequency domain solutions satisfy the boundary conditions (noting they must satisfy the Helmholtz equation by construction).
		
		\subsection{Continental shelf and slope}
\begin{figure}
			\centering    
			(a)\hfill
			\begin{subfigure}{}
				\includegraphics[width=0.98\textwidth]{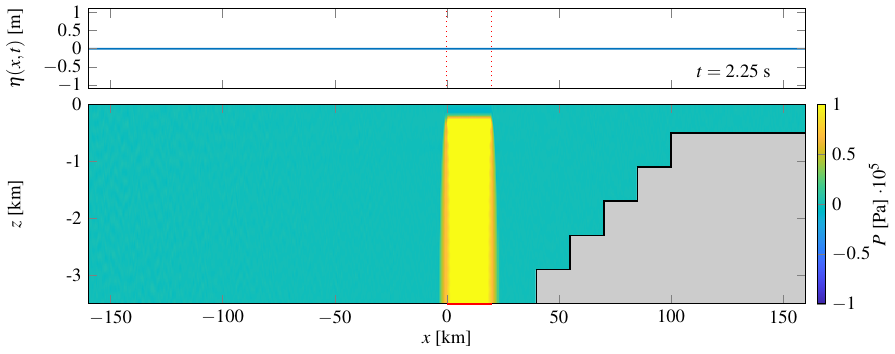}
			\end{subfigure}\\
			(b)\hfill
			\begin{subfigure}{}
				\includegraphics[width=0.98\textwidth]{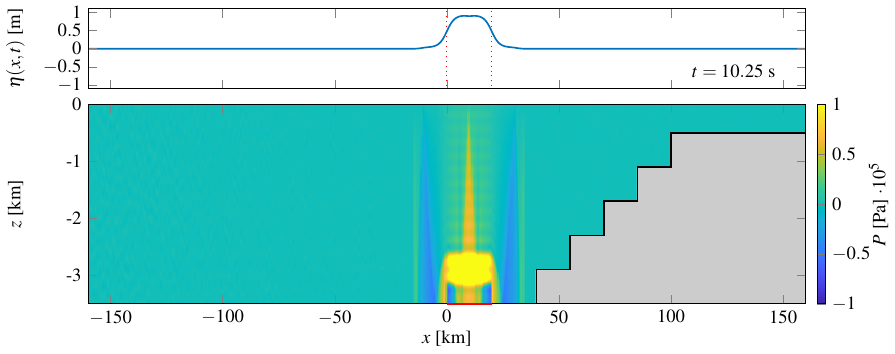}
			\end{subfigure}\\
			(c)\hfill
			\begin{subfigure}{}
				\includegraphics[width=0.98\textwidth]{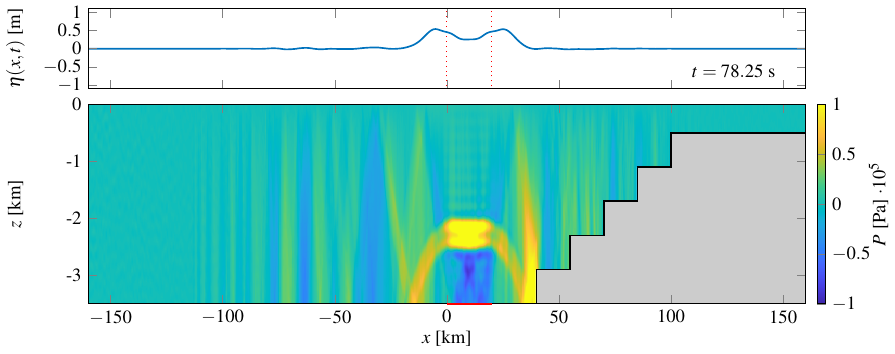}
			\end{subfigure}\\
			\caption{Compression waves near a continental shelf caused by a constant velocity rupture (Movie $1$ shows full animation). Each time panel presents the free-surface profile above a heat-map of the dynamic pressure in the fluid. The red dotted lines in the free-surface plot indicated the edge of the rupture zone. The rupture surface is indicated in the pressure plot by the red line from $0<x<20$.
			}
			\label{fig:9}
		\end{figure}

		In this subsection, we show the surface profile and pressure distribution inside the water column for waves generated near continental shelf- and slope-type bathymetry. Our continental shelf is characterised by a maximum depth of $3.5$km and a minimum depth of $500$m. The change in depth occurs over five equally spaced steps from $x=40$km to $x=100$km (Fig.~\ref{fig:9}), leading to an average inclination of approximately $2.3^\circ$. The region $x\in[0,20]$km is chosen as our rupture zone.
		
		\subsubsection{Constant rupture velocity}\label{sec:shelf_const}
		
\begin{figure}[h]
			\centering
			(a)\hfill
			\begin{subfigure}{}
				\includegraphics[width=0.98\textwidth]{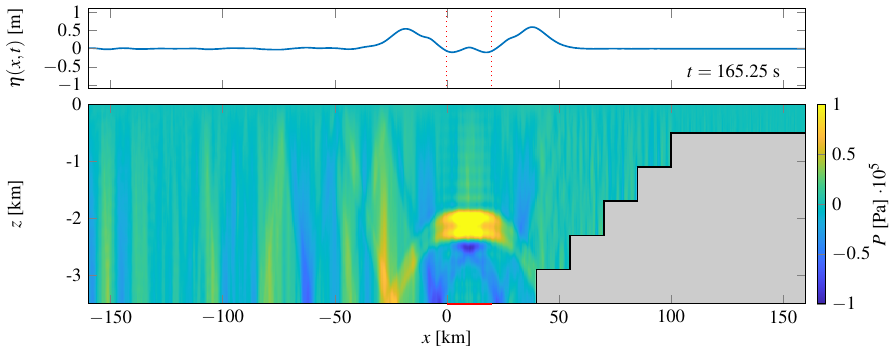}
			\end{subfigure}\\
			(b)\hfill
			\begin{subfigure}{}
				\includegraphics[width=0.98\textwidth]{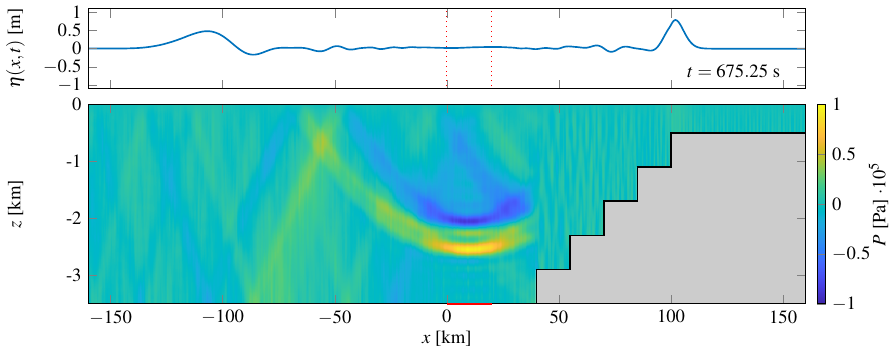}
			\end{subfigure}\\
			\caption{Same as Fig. \ref{fig:9} at later time points. Movie $2$ shows full animation.
			}
			\label{fig:9b}
		\end{figure}

		For a constant velocity rupture, the evolution of the surface wave profile is shown in Figs. \ref{fig:9} and \ref{fig:9b} with snapshots at $t=2.25,\,10.25,\,78.25,\,165.25$, and $675.25$s. In the first snapshot, we see a column of positive pressure rising from the rupture surface, travelling at the speed of sound, driven by the constant-speed movement of the boundary. This pressure column will reflect off of the water's surface, changing its sign before continuing back down to the seabed and reflecting again without changing its sign. As this process repeats, the pressure column makes four trips between the seabed and the surface such that it is positive and moving up when the rupture zone stops moving. The combined interference of this pressure column results in a localised positive pressure that will continue to bounce between the seabed and the surface (Fig. \ref{fig:9}(b)).
		
		As time continues, pressure waves peel off from the localised pressure pulse and propagate away from the rupture region (Fig.\ref{fig:9}(c)). These pressure waves slowly align themselves with the vertical axis as they propagate away from the continental shelf, in the constant depth region, because the finite depth domain acts as a waveguide. The pressure waves travelling towards the continental shelf scatter at the depth changes and result in a higher frequency pressure pattern over the shallower depth regions (Fig. \ref{fig:9b}).
		
		Focussing on the free-surface profiles, we can see that the effect of compressibility means that the surface only begins to move noticeably after the pressure column has had time to reach the surface (Fig. \ref{fig:9}).  Once the surface starts to move, it lifts above the rupture (Fig. \ref{fig:9} (b)) before splitting into two gravity waves moving in both directions, which is consistent with early-time symmetric solutions described in the author's previous works \citep{Das2024,Pethiyagoda_Das_Meylan_2024}. Comparing the two gravity waves at a later time point (Fig. \ref{fig:9b}), we can see that the decrease in depth on the right has caused the travelling wave to become steeper. Additionally, the previously mentioned pressure waves generated the AGWs, which appear as small ripples on the surface. Movie $1$ and Movie $2$ show the simulation videos for shorter and longer time scales. 
		
		\subsubsection{Parabolic rupture velocity}\label{sec:shelf_para}
		\begin{figure}
			\centering    
			(a)\hfill
			\begin{subfigure}{}
				\includegraphics[width=0.98\textwidth]{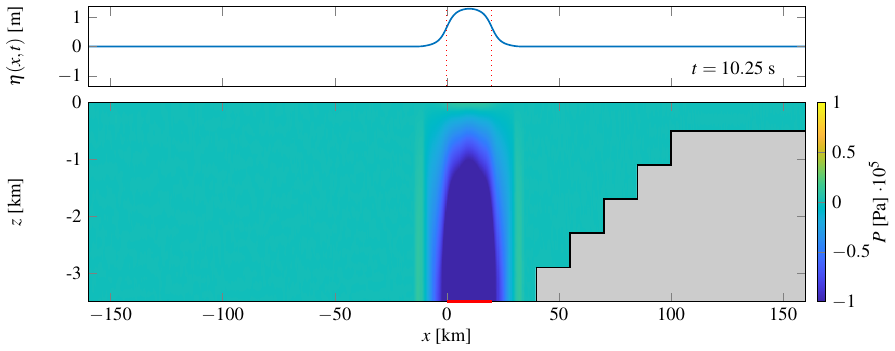}
			\end{subfigure}\\
			(b)\hfill
			\begin{subfigure}{}
				\includegraphics[width=0.98\textwidth]{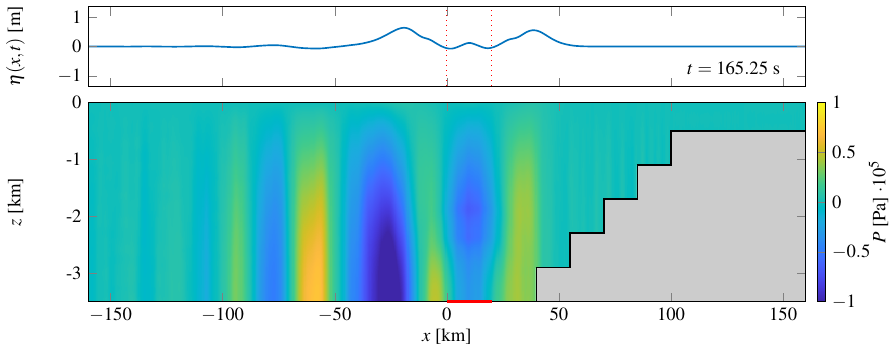}
			\end{subfigure}\\
			(c)\hfill
			\begin{subfigure}{}
				\includegraphics[width=0.98\textwidth]{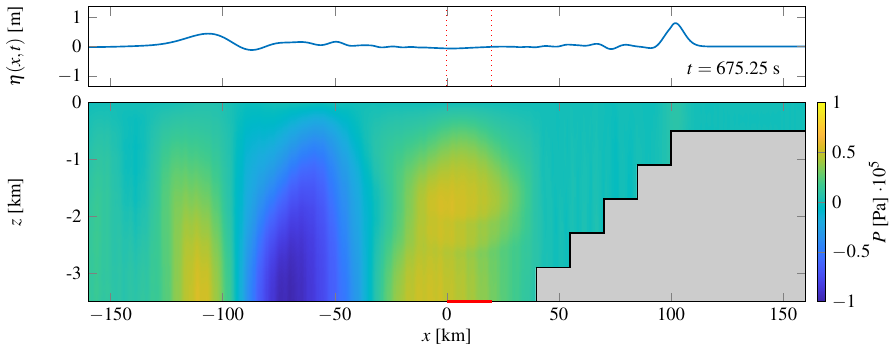}
			\end{subfigure}\\
			\caption{Compression waves near a continental shelf caused by a parabolic rupture velocity (check Movie $3$ for the full animation). Each time panel presents the free-surface profile above a heat-map of the dynamic pressure in the fluid. The red dotted lines in the free-surface plot indicate the edge of the rupture zone. The rupture surface is indicated in the pressure plot by the red line from $0<x<20$.
			}
			\label{fig:conshelf_para}
		\end{figure}
		
		The solutions for a parabolic rupture velocity are shown in Fig. \ref{fig:conshelf_para}  with snapshots at $t=10.25,\,165.25,$ and $675.25$s. The dynamic pressure above the rupture exhibits a more non-localised profile in the vertical direction compared with the previous constant velocity rupture solutions (Fig. \ref{fig:9b}). This profile leads to a more defined banded structure in the pressure waves propagating to the left away from the rupture (Fig. \ref{fig:conshelf_para}b). As a consequence, the AGWs on the surface are of a greater magnitude than what was previously seen for the constant velocity rupture.
		
		The initial banded structure (i.e., low frequency in the $z$-direction) of the right propagating pressure waves lead to greater reflection off of the continental shelf and, therefore, significantly smaller AGWs appearing on the continental shelf when compared to the constant velocity rupture solutions. Additionally, the reflected pressure waves interact with original left propagating waves, leading to much larger AGWs in both wavelength and amplitude compared with the early time solutions (Fig. \ref{fig:conshelf_para}c). The Movie $3$ shows the simulation video.
		
		\subsection{Trench and ridge}
		In this subsection, we show the surface profile and dynamic pressure inside the water column for waves generated near an underwater trench mountain ridge. The bathymetry in kilometres is taken as a piecewise constant approximation of 
		\[
		h(x)=-2.5-6(x-20)(x-80)H(x-20)H(x-80)+6(x-120)(x-180)H(x-120)H(x-180).
		\]
		The parabolic arcs that define the trench and ridge are divided into five levels (Fig.~\ref{fig:trench_ridge_const}). The region $x\in[-10,10]$km is chosen as our rupture zone.
		
		\subsubsection{Constant rupture velocity}\label{sec:trench_ridge_const}
		\begin{figure}
			\centering    
			(a)\hfill
			\begin{subfigure}{}
				\includegraphics[width=0.98\textwidth]{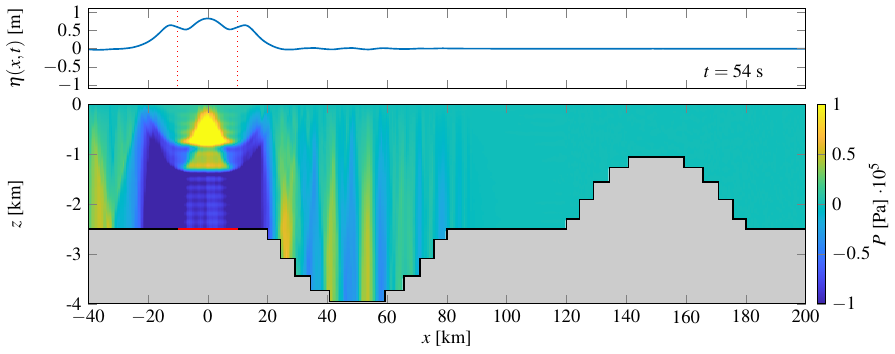}
			\end{subfigure}\\
			(b)\hfill
			\begin{subfigure}{}
				\includegraphics[width=0.98\textwidth]{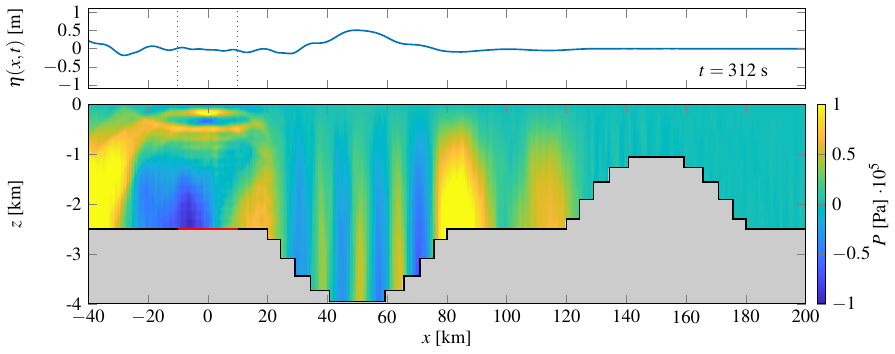}
			\end{subfigure}\\
			(c)\hfill
			\begin{subfigure}{}
				\includegraphics[width=0.98\textwidth]{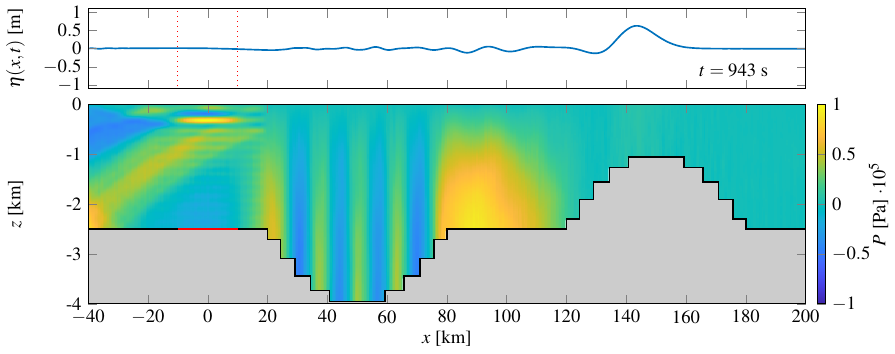}
			\end{subfigure}\\
			\caption{Compression waves near an underwater trench and a mountain ridge caused by a constant velocity rupture. Each time panel presents the free-surface profile above a heat-map of the dynamic pressure in the fluid. The red dotted lines in the free-surface plot indicate the edge of the rupture zone. The rupture surface is indicated in the pressure plot by the red line from $-10<x<10$. Movie $4$(a) and Movie $4$(b) show full animations.
			}
			\label{fig:trench_ridge_const}
		\end{figure}
		
		Figure \ref{fig:trench_ridge_const} provides snapshots of the surface displacement and dynamic pressure caused by a constant velocity rupture for $t=54,\,312$, and $943$s. The first thing to address is that even though the rupture speed is the same as that presented in \S\ref{sec:shelf_const}, the shallower rupture location has resulted in different interference patterns in the pressure and, therefore, a less localised pressure pulse bouncing between the surface and the bottom. This change in local pressure behaviour has the effect of altering the pressure waves that propagate away from the rupture region, leading them to appear more banded than those observed in Fig. \ref{fig:9} but still less so than those created by the parabolic rupture velocity (Fig. \ref{fig:conshelf_para}).
		
		As with previous examples, pressure waves begin propagating away from the rupture location. These waves move over the trench with some reflection occurring as the waves leave the trench. These reflections begin to excite a standing wave pattern within the trench. As the waves move past the trench, they interact with and mostly reflect back off of the ridge, much like the waves in \S\ref{sec:shelf_para}. These additional reflected waves from the ridge return past the trench and further excite the standing waves within the trench such that they are readily apparent on the free-surface (Fig. \ref{fig:trench_ridge_const}c).
		
		The gravity waves generated by the rupture behave in much the same way as the previous examples. The gravity waves propagate away from the rupture area, maintaining the same qualitative shape while slightly broadening as they pass over the trench and steepening as they pass over the ridge. The simulations are provided as Movie $4$(a) and Movie $4$(b).
		
		\begin{figure}
			\centering    
			(a)\hfill
			\begin{subfigure}{}
				\includegraphics[width=0.98\textwidth]{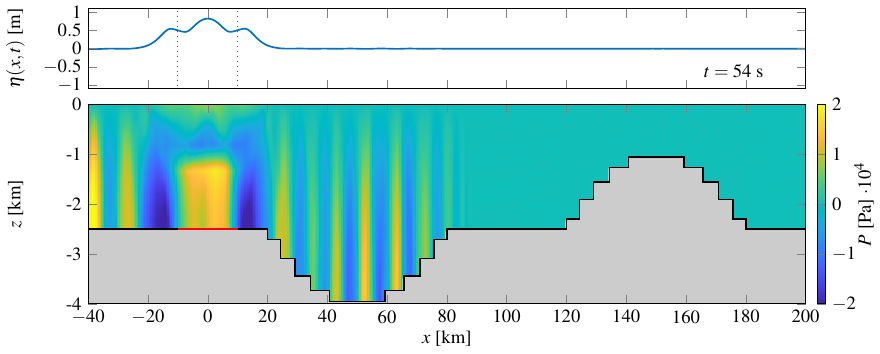}
			\end{subfigure}\\
			(b)\hfill
			\begin{subfigure}{}
				\includegraphics[width=0.98\textwidth]{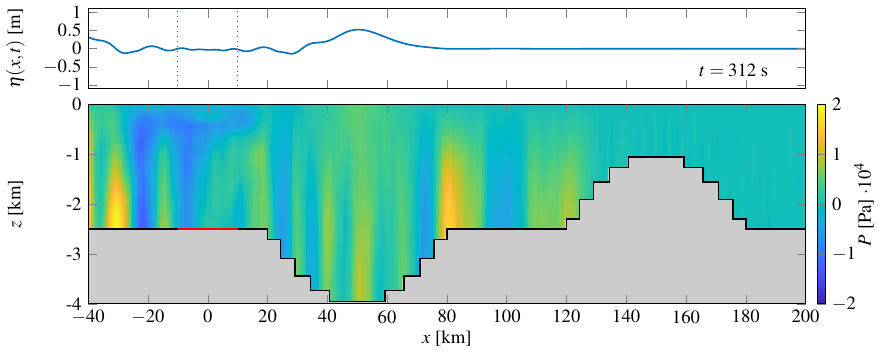}
			\end{subfigure}\\
			(c)\hfill
			\begin{subfigure}{}
				\includegraphics[width=0.98\textwidth]{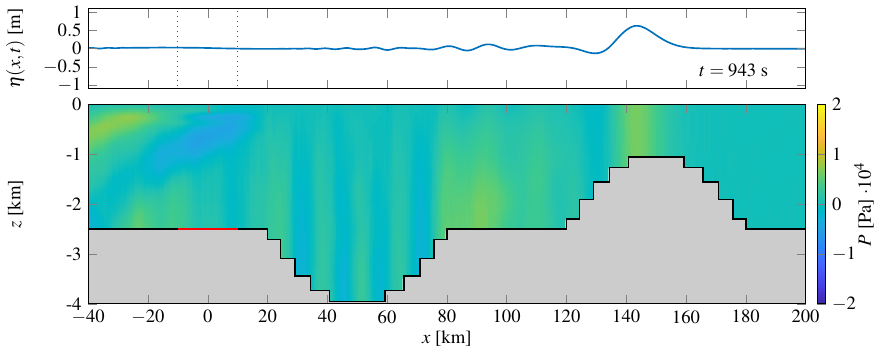}
			\end{subfigure}\\
			\caption{Compression waves near an underwater trench and a mountain ridge caused by a parabolic rupture velocity. Each time panel presents the free-surface profile above a heat-map of the dynamic pressure in the fluid. The red dotted lines in the free-surface plot indicate the edge of the rupture zone. The rupture surface is indicated in the pressure plot by the red line from $-10<x<10$. Movie $5$(a) and Movie $5$(b) show full animations.
			}
			\label{fig:trench_ridge_para}
		\end{figure}
		\subsubsection{Parabolic rupture velocity}
		
		As with the previous example, the shallower depth has changed how the pressure above the rupture appears compared to the deeper rupture location \S\ref{sec:shelf_para}. In fact, this depth has resulted in a dynamic pressure profile that is almost an order of magnitude smaller (note the colour bar in Fig.~\ref{fig:trench_ridge_para} compared with Figs.~\ref{fig:9}-\ref{fig:trench_ridge_const}). The decreased pressure means the surface gravity waves dominate the wave profile.
		
		Comparing with \S\ref{sec:trench_ridge_const}, we see that the change in rupture speed has resulted in the propagating pressure waves having a smaller wavelength in the $x$-direction. We believe this property leads to the pressure waves more easily passing over the trench region, reducing the excitation of the standing waves within the trench (Fig.~\ref{fig:trench_ridge_para}b). The pressure waves that are reflected off the ridge structure return to the trench and further excite the standing waves, much like the results of constant rupture velocity. Full animations are provided as Movie $5$(a) and Movie $5$(b).
		
		\section{Conclusion}\label{sec:conclusion}
		In this paper, we have created a general formulation for the calculation of fluid pressure and free-surface displacement caused by vertical motion of the ocean floor with an arbitrary piecewise constant bathymetry. We include slight ocean compressibility, including static compression, to facilitate the modelling of acoustic gravity waves. The propagation of AGWs is highly dependent on the bathymetry, and therefore, the generality of our formulation allows for the straightforward implementation and study of different bathymetric features, such as a continental shelf, underwater trenches, and mountain ridges, or any combination of the three. We have shown solutions for bathymetric profiles of a continental shelf, and an underwater trench followed by a ridge, with an ocean bed rupture moving vertically at either constant speed or parabolic speed. With just the four cases studied in this paper, we have observed a large variance in pressure wave behaviour in the rupture zone, indicating that the rupture depth and velocity profile must be accurately captured for the appropriate conclusion to be drawn from the simulation results. Over all our solutions, we have observed that propagating internal pressure waves that appear as vertical bands lead to more reflected waves by any bathymetry rise (e.g., like a continental shelf or ridge). The strong dependence of the pressure profiles on the rupture behaviour and the bathymetry begs the question: how much information can be inferred from measured underwater pressure? 
		
		This formulation for an arbitrary depth can be used to expand the results of \cite{Gomez_Kadri_2021b,Gomez_Kadri_2021a}, who used a simplified model ignoring the effects of gravity, to a broader scope. The simplicity of the physical setup poses limitations in capturing the effects of ocean bed properties like elasticity, porosity, and viscoelasticity, and the effect of variable sound speed across the water column. However, the computation accuracy of the present method motivates us to incorporate these effects in future research and provide a more holistic approach.
		
		\section*{Supplementary material}
		The animation movies are uploaded as supplementary material (doi: 10.6084/m9.figshare.30226018).
		\section*{Acknowledgement}
		SD acknowledges the University of Newcastle, Australia, for hosting him as SIRE Fellow, during which the work was initiated.
		
		\section*{Funding}
		This work was partially supported by the Science and Engineering Research Board (SERB), Govt. of India, through the SERB International Research Experience (SIRE) fellowship (Sanction No. SIR/2022/000994) awarded to SD. 
		
		\section*{Declaration of interests}
		The authors report no conflict of interest.
		
		\section*{DATA AVAILABILITY}
		Data sharing does not apply to this article as no new data were created or analyzed in this study. The computer code used to generate the solutions is available from the corresponding author upon reasonable request.
		
		\providecommand{\noopsort}[1]{}\providecommand{\singleletter}[1]{#1}%

		\appendix
		\section{Convergence}
		
\begin{figure}
			\centering
			(a) $N=15$\hfill
			\begin{subfigure}{}
				\includegraphics[width=0.98\textwidth]{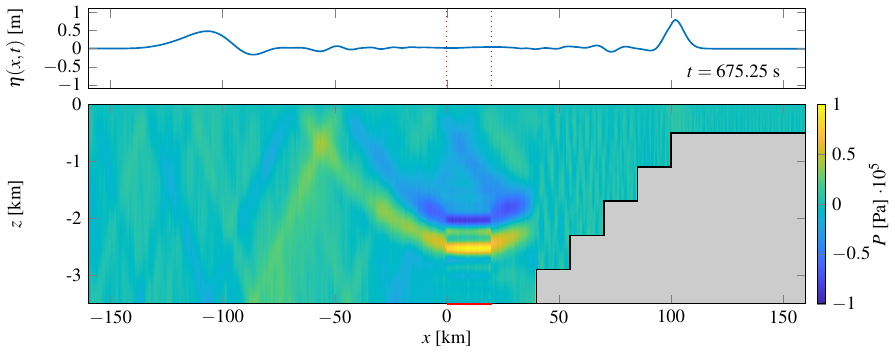}
			\end{subfigure}\\
			(b) $N=30$\hfill
			\begin{subfigure}{}
				\includegraphics[width=0.98\textwidth]{continental_shelf_t=675.25.pdf}
			\end{subfigure}\\
			(c) $N=60$\hfill
			\begin{subfigure}{}
				\includegraphics[width=0.98\textwidth]{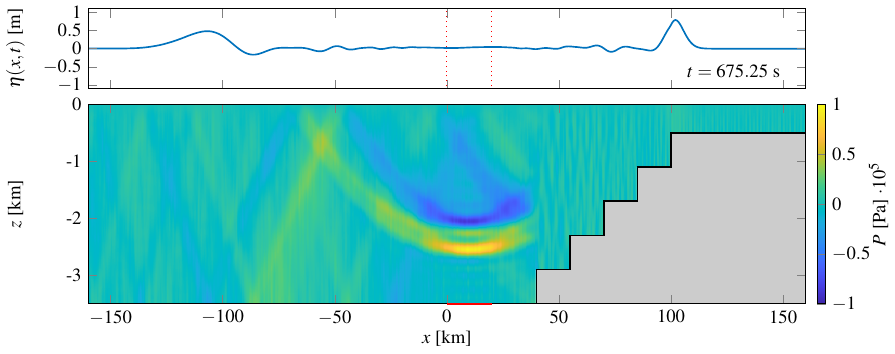}
			\end{subfigure}\\
			\caption{Same as Fig. \ref{fig:9} at $t=675.25$ for three difference choices of $N$. (a) $N=15$, (b) $N=30$, and (c) $N=60$.
			}
			\label{fig:10}
		\end{figure}

		To make a remark about the convergence of the mode matching scheme, we present here the solution to the continental shelf bathymetry in Fig. \ref{fig:10} using three different series truncation values $N=15$, $N=30$ (used for the rest of the paper), and $N=60$. We can see that using only fifteen AGW modes does not give a smooth solution near the rupture location; however, $N=30$ and $N=60$ give smooth, visually identical results. We note that away from the rupture location, the $N=15$ solution is very similar to the other solutions.
		
	\end{document}